\documentclass[a4paper]{jpconf}
\usepackage[mathcal]{euscript}
\usepackage[latin1]{inputenc}
\usepackage{latexsym}
\usepackage{amsmath}
\usepackage{hyperref}
\usepackage{amsmath}    
\usepackage{graphicx}   
\usepackage{verbatim}  
\usepackage{color}      
\usepackage{subfigure}  
\usepackage{hyperref}   
\usepackage{bm}
\usepackage{graphicx}
\usepackage{amssymb}
\usepackage{bbm}
\usepackage{float}
\usepackage{slashed}
\usepackage{amsfonts}
\usepackage{euscript}
\usepackage{amsmath}    
\usepackage{mathrsfs} 
\usepackage{bbding}
\usepackage{pifont}
\usepackage{cancel}
\begin{document}

\title{Some not-so-common ideas about gravity}

\author{Ra\'ul Carballo-Rubio$^{1}$, Carlos Barcel\'o$^1$ and Luis J Garay$^{2,3}$}
\address{$^1$ Instituto de Astrof\'isica de Andaluc\'ia (IAA-CSIC), Glorieta de la Astronom\'ia, 18008 Granada, Spain}
\address{$^2$ Departamento de F\'{\i}sica Te\'orica II, Universidad Complutense de Madrid, 28040 Madrid, Spain}
\address{$^3$ Instituto de Estructura de la Materia (IEM-CSIC), Serrano 121, 28006 Madrid, Spain}
\ead{raulc@iaa.es}

\begin{abstract}{
Most of the approaches to the construction of a theory of quantum gravity share some principles which do not have specific experimental support up to date. Two of these principles are relevant for our discussion: (i) the gravitational field should have a quantum description in certain regime, and (ii) any theory of gravity containing general relativity should be relational. We study in general terms the possible implications of assuming deviations from these principles, their compatibility with current experimental knowledge, and how can they affect future experiments.
}
\end{abstract}

%
%

\section{Introduction}

The acceptance of certain principles or axioms is a necessary scientific practice. In natural sciences, these principles should have a strong experimental corroboration in the best of the cases or, at least, do not contradict any experimental fact. When a theoretical paradigm shows signs of needing a revision, either because of internal or external (when compared with experiments) inconsistencies, these principles which lack of any direct experimental support are usually the first suspects in order to be replaced.

There is accumulative evidence that general relativity must be completed at high energies by a yet unknown theory (see, e.g., \cite{Wald2010} and references therein). This observation follows from purely theoretical considerations, as there is no experiment up to date which is not consistent with the theory \cite{Will2006}. Theoretical investigations of the high-energy properties of the gravitational interaction routinely assume certain principles as valid (see, e.g., \cite{Rovelli2004}). Of particular interest for us are the following two: (i) the gravitational field should have a quantum description in certain regime, and (ii) any theory of gravity containing general relativity should be relational. In this contribution we emphasize that there exist alternatives to these principles which are compatible with current experimental knowledge. Not only this but, if these ideas are taken seriously, they can lead to specific experimental signatures. Personally, I think that diversity is beneficial (not only for science) but, moreover, that it is worth it to understand the possible implications of deviations from these principles in order to build empirical confidence on them.

It is our aim to explore two ideas, in principle independent, although they could present interesting interconnections. The first idea is that some properties or elements of the gravitational interaction could be essentially classical in nature. This idea has been proposed by many people at different times; here we just focus in some general conceptual issues which appear to be model-independent. The second idea is that a convenient theory of gravity do not need to be relational, in the sense that it could contain non-dynamical structures such as a preferred volume element or a trivial (i.e., with no horizons) non-dynamical causal structure.

Our goal is to understand the possible consequences of these ideas. We will see that, if gravity has to be essentially classical, then one should expect deviations from standard quantum mechanics in mesoscopic experiments involving gravity and matter fields. On the other hand, if one has a non-relational theory of gravity, one should expect deviations from the usual gravitational phenomenology. In particular, a fixed volume element may permit to avoid the cosmological constant problem, while having a trivial non-dynamical underlying causality makes possible the construction of totally different stories for the gravitational collapse process.

\section{Hybrid classical-quantum theories}

In the first part we are going to study a specific model for a general classical-quantum hybrid system. The possibility of formulating a consistent hybrid classical-quantum theory has been approached in many different ways. In all these approaches one starts with initially separated purely classical and quantum sectors and then makes them interact in order to analyze the outcome. Without pretending to be exhaustive, we can classify these approaches in the following categories: (1) approaches that try to maintain the use of quantum states (or density matrices) to describe the quantum sector and trajectories for the classical sector~\cite{BoucherTraschen1988,Anderson1995}, (2) those that first formulate the classical sector as a quantum theory~\cite{Koopman1931,Neumann1932,Sudarshan1976} and then work with a  formally  completely quantum system~\cite{Sherry1978,Sherry1979,Gautam1979,PeresTerno2001,Terno2006,Buric2012,Buric2013,Buric2013b,Buric2014}, (3) conversely, those that first formulate the quantum sector as a classical theory~\cite{Heslot1985} and then work with a formally completely classical system~\cite{Elze2011,Alonso2010,Alonso2011,Alonso2012}, and (4) approaches that take the quantum and the classical sectors to a common language and then extend it to a single framework in the presence of interactions, for instance, using Hamilton-Jacobi statistical theory for the classical sector and Madelung representation for the quantum sector~\cite{HallReginatto2005,Hall2008,Chua2012} or modeling classical and quantum dynamics starting from Ehrenfest equations~\cite{Bondar2012}. This classification  is not sharp and in some cases it is subject to interpretation, but it may be useful as a way to organize the possible procedures and conceptual viewpoints in the enterprise of  constructing a hybrid theory. For more details on issues with these formulations, see \cite{Barcelo2012,Elze2013} and references therein.

Approaches in category 2 deal with a perfectly defined quantum theory based on Koopman-von Neumann-Sudarshan~\cite{Koopman1931,Neumann1932,Sudarshan1976} translation of classical mechanics into the language of Hilbert spaces. Within this approach, the work of Peres-Terno is often mentioned as an important objection to the consistency of hybrid configurations. They proved that, in the presence of interaction, Heisenberg's equations of motion of the canonical-variable operators (Heisenberg picture) are necessarily different from what would have been obtained by assuming the two sectors to be quantum. From this they conclude (as we will see, somewhat precipitately) that the classical (Ehrenfest) limit of the classical-quantum (CQ) system will be necessarily different from that obtained directly from the corresponding quantum-quantum (QQ) system. On the other hand, they argue that these two limits should coincide, that is, that any hybrid theory should comply with this {\em definitive benchmark}, as they call it. Then, they finally conclude that hybrid systems, at least of this kind, are unphysical.

Here, first of all, we want to stress that this benchmark only makes  sense if one tacitly assumes that a hybrid system should be just some suitable limit of a purely quantum system. If this is not assumed, we can invert the logic and see the failure to satisfy this condition as a clear sign of new physics in the CQ interplay. Second, as we will see the Peres-Terno model has many interesting features, being more complex than just exposed.  When looking at the quadratic Peres-Terno model in more detail, one realizes that {\em it can} be made to comply with the above benchmark by just adding some constraints to the system. In fact, to add constraints to the system is more than reasonable because its very formulation incorporates more degrees of freedom than the QQ theory with which it is compared. Nonetheless, in a second  twist, we have shown that this hybrid system still exhibits some sort of refined Peres-Terno no-go result: One cannot require that the dynamical equations for the second-order momenta (i.e., the dispersions) be  equal to those in the QQ system (which for these system are equal to the CC ones). Going beyond quadratic interactions we also show that there is no  way in general to fulfill the Peres-Terno benchmark by adding constraints. Overall, we are led to conclude that either the evolution of the expectation values or the evolution of the dispersions will exhibit some new physics beyond both QQ and CC systems. 

The starting point of this particular hybrid scheme is the Koopman-von Neumann-Sudarshan formulation of classical mechanics~\cite{Koopman1931,Neumann1932,Sudarshan1976,Mauro2003}. This formulation translates the usual description of classical mechanics in terms of symplectic manifolds to a quantum-mechanical language by associating to each physical observable  a self-adjoint operator in a suitable Hilbert space and by implementing the time evolution as a unitary operator. Once the classical sector is treated formally as quantum, one can describe a CQ interaction by using the tensor product of Hilbert spaces, as in a pure quantum-mechanical theory~\cite{Ballentinebook}. It is in developing this program that one has to face a number of difficulties.

To make the discussion self-contained let us briefly summarize the 
Koopman-von Neumann-Sudarshan formalism. For the sake of simplicity we shall deal with one-dimensional systems although the discussion can be easily generalized to an arbitrary number of degrees of freedom. The quantum sector will be represented by a Hilbert space $\mathcal{H}_{\mbox{\tiny Q}}$ with the standard position and momentum operators,
\begin{equation}
[\hat{x},\hat{k}]=i\hbar.
\end{equation}
In the classical sector we  will also have position and momentum operators  
defined over a Hilbert space $\mathcal{H}_{\mbox{\tiny C}}$, but in this case they commute,
\begin{equation}
[\hat{q},\hat{p}]=0.
\label{eq:clasop}
\end{equation}
The elements of the pre-Hilbert space $\overline{\mathcal{H}}_{\mbox{\tiny C}}$ are  taken to be the ``classical'' wave functions $\psi(q,p)$, whose square equal the classical distribution functions:  
\begin{equation}
 |\psi|^2=\rho_{\mbox{\tiny C}}(q,p).
\label{eq:claswave}
\end{equation}%
The action of the classical operators (\ref{eq:clasop}) is then multiplicative,
\begin{equation}
\hat{q}\,\psi(q,p)=q\psi(q,p),\qquad \hat{p}\,\psi(q,p)=p\psi(q,p).
\end{equation}
As is well known from classical statistical mechanics, the evolution of a classical probability distribution $\rho_{\mbox{\tiny C}}(q,p)$ is given by the Liouville equation
$\partial_t \rho_{\mbox{\tiny C}} = \hat L \rho_{\mbox{\tiny C}}$~\cite{Ballentinebook}.  Since the Liouville operator $\hat{L}$ is linear in the derivatives, the evolution equation for $\psi$ is the same as that for $\rho_{\mbox{\tiny C}}$:
\begin{align}
i\hbar \partial_t \psi=&{} \hat{H}_{\mbox{\tiny CL}} \psi,
\qquad \hat H_{\mbox{\tiny CL}}:=  i\hbar \hat L= \widehat{\partial_p H_{\mbox{\tiny C}}}~\hat{p}_q + \widehat{\partial_q H_{\mbox{\tiny C}}}~\hat{q}_p.
\end{align}
Here $H_{\mbox{\tiny C}}(q,p)$ is the classical Hamiltonian and we have defined the operators $(\hat p_q,\hat q_p)$ to be  canonically conjugate  to   $(\hat{q},\hat{p})$:
\begin{equation}
[\hat{q},\hat{p}_q]=i\hbar=[\hat{q}_p,\hat{p}],\qquad [\hat{q},\hat{q}_p]=[\hat{p},\hat{p}_q]=0;
\end{equation}
which are correspondingly represented by
\begin{equation}
\hat{p}_q=-i\hbar \partial_q,\qquad\hat{q}_p=i\hbar \partial_p.
\end{equation}
In the following these variables will be called \emph{unobservable} variables.

Moreover, under reasonable assumptions about the  classical  Hamiltonian, the Liouville Hamiltonian $\hat{H}_{\mbox{\tiny CL}}$ is   essentially self-adjoint   in the inner product,
\begin{equation}
(\psi,\phi):=\int \text{d}q\text{d}p\,\psi^*\phi,\qquad\psi,\phi\in\overline{\mathcal{H}}_{\mbox{\tiny C}},
\label{eq:inner}
\end{equation}
so it generates a unitary evolution~\cite{ReedSimon1981}. The completion of $\overline{\mathcal{H}}_{\mbox{\tiny C}}$ in the inner product (\ref{eq:inner}) constitutes the classical Hilbert space $\mathcal{H}_{\mbox{\tiny C}}$.

Hereafter let us work for convenience in the Heisenberg picture in which the operators (\ref{eq:clasop}) carry the time dependence. With the previous definitions it is direct to check that the Heisenberg equations for the operators $(\hat{q},\hat{p})$ have the same form as the classical Hamilton equations. The unobservable variables $(\hat{q}_p,\hat{p}_q)$ have their own evolution without influencing the physical sector $(\hat{q},\hat{p})$. So, at least at this level, their evolution is irrelevant. Let us point out that, concerning the classical observable sector and directly related to this decoupling, there exists a symmetry under phase  transformations of the form
\begin{equation}
\psi(q,p)\longrightarrow e^{i\varphi(q,p)}\psi(q,p).
\label{eq:gauge}
\end{equation}
As we will see, this ambiguity disappears when developing a CQ interaction scheme. To end this summary, let us mention that it can be rigorously proved~\cite{Koopman1931,Neumann1932} that the Koopman-von Neumann-Sudarshan formulation of classical mechanics is equivalent to the standard one in terms of symplectic manifolds and permits the use of operator techniques to treat classical problems. In particular, it is quite useful to study some aspects of statistical mechanics and ergodic theory~\cite{ReedSimon1981}.

Let us now consider an interacting CQ system~\cite{Sudarshan1976,PeresTerno2001}. The space of states is the tensor product $\mathcal{H}_{\mbox{\tiny C}} \otimes\mathcal{H}_{\mbox{\tiny Q}}$, and the total Hamiltonian operator is
\begin{equation}
\hat{H}_{\mbox{\tiny T}}=\hat{H}_{\mbox{\tiny CL}}+\hat{H}_{\mbox{\tiny Q}}+
\hat{H}_{\mbox{\tiny I}},
\end{equation}
where $\hat{H}_{\mbox{\tiny Q}}$ stands for the Hamiltonian of the quantum system and $\hat{H}_{\mbox{\tiny I}}$ provides the CQ interaction. At this point, the first question with a non-straightforward answer appears: How do we determine the form of the operator $\hat{H}_{\mbox{\tiny I}}$ from its classical counterpart $H_{\mbox{\tiny I}}$?

Given any interaction term, one can easily check that the classical Hamiltonian equations of motion (considering the system completely classical) and quantum Heisenberg equations are formally equivalent. The proposal of Peres and Terno~\cite{PeresTerno2001} as the definite benchmark for an acceptable classical-quantum hybrid was that any hybridization of this system should respect this equivalence. Then, the result which they proved is that even for this simple system there is no term $\hat{H}_{\mbox{\tiny I}}=\hat{H}_{\mbox{\tiny I}}(\hat{q},\hat{p},\hat{x},\hat{k},\hat{q}_p,\hat{p}_q)$ which fulfills this condition, so they rejected this kind of hybrid CQ dynamics for not being physically meaningful.

In brief, the problem emerges as follows. If one wants the operators of the quantum sector $(\hat{x},\hat{k})$ to appear in the equations of motion of the classical variables $(\hat{q},\hat{p})$, one has to introduce the unobservable variables $(\hat{q}_p,\hat{p}_q)$ in the interaction term. But, by doing that, the decoupling of the unobservable sector no longer holds. That is, the equations of motion of the variables in the physical sector will contain the unobservable operators explicitly. As the unobservable operators do not appear in the QQ theory, the equations of motion of the CQ and QQ theories are different. Notice that it is the description of the effect of the quantum variables on the classical variables which leads to these features, which goes further than strict semiclassical physics, i.e., quantum mechanics in presence of classical fields. At this point it is interesting to have in mind that experiments have explored so far the semiclassical realm, so that we have no empirical clue about the behavior of physical systems near the classical-quantum interface.  

With the risk of overinterpreting Peres and Terno's logic, in our view they took this condition as a definite benchmark because they identified formally having  equal Heisenberg equations with obtaining an appropriate correspondence principle. However these are, in principle, logically distinct issues, as we have shown \cite{Barcelo2012}. Moreover, this benchmark is fully appropriate if one assumes that the hybrid theory is just a particular approximation of the straight QQ theory (here by straight we mean the standard quantization one would have performed to the classical Hamiltonian for two interacting harmonic oscillators). However, if one is looking at hybrid systems as examples of new physics, then the violation of this benchmark could be interpreted positively. The precise prescription of an interaction term $\hat{H}_{\mbox{\tiny I}}$ should take into account additional physical insights coming from the detailed characteristics of the variable that is being regarded as classical, and how it is affected by the interaction with quantum systems.    

All in all, our analysis shows that one can formulate consistent hybrid theories, but these theories incorporate new physics. They are not suitable limits of a direct quantization of the entire system. We should distinguish this straight quantum theory from the possibility of using a quantum formalism to describe all the physics of the hybrid system, as happens in the Koopman-von Neumann-Sudarshan formalism. Among other things, the new physics appears to bring about new degrees of freedom, which should be associated with the additional level of description of the quasi-classical (macroscopic) variables one would need to prescribe the precise form of the quantum backreaction to these variables. In this respect, the measurement process as presented by the Copenhagen interpretation could be seen as an extreme case of interaction between a classical and a quantum system in which there is no transfer of fluctuations into the classical variables. The measurement theory of Sudarshan \cite{Sudarshan1976,Sherry1978,Sherry1979} goes one step further, allowing some transfer of fluctuations while they do not compromise the classical integrity of the variables. Another interesting lesson is that in a hybrid context pure notions such as quantum or classical are only limiting notions, which are very useful in isolated situations but are, strictly speaking, nonexistent; in a hybrid setting everything might have a certain degree of ``quantumness'' (classicality).

At the end of the day, only new experiments in the mesoscopic realm can help us decide whether in the behavior of a complete system there is something beyond quantum dynamics of microscopic quantum constituents. Proposals such that those of Di\'osi and Penrose~\cite{Diosi1987,Penrose1996,Marshalletal2003} or Guirardi et al.~\cite{Ghirardietal1985} of new physics at the classical-quantum cut find additional justification in light of these hybrid models. In particular, the proposal that gravity, with its stubborn resistance to quantization, might play a role in going beyond standard quantum mechanics is a powerful idea that we should not forsake.

\section{Non-geometric gravity}

The geometric vision of gravity embodied in general relativity has been highly successful up to date in the explanation of gravitational phenomena. However, as discussed above, general relativity should be replaced at high energies. While the majority of approaches to construct a theory of quantum gravity consider geometry as a fundamental tool to describe the high-energy regime, emergent gravity approaches show that non-relational structures could play an important role. It is important to notice that, within this context, these non-relational properties are not regarded as fundamental, but rather as being also emergent.

\subsection{Volume element}

One of the constructive processes which can be used to reach the theory of general relativity is the self-coupling procedure of gravitons (see~\cite{Ortin2004,Barcelo2014} for a thorough discussion of the history of the subject and the different assumptions needed to obtain the result). It is commonly said that general relativity is the only solution to the self-coupling problem \cite{Deser1970}. However, it has been stated recently \cite{Barcelo2014,BCRG2014} that there exists another solution which is not structurally equivalent to general relativity, but it is described by the structure of Weyl transverse gravity \cite{Alvarez2006,Blas2007,Alvarez2010,Alvarez2012}. Most importantly for our purposes, the resulting nonlinear theory displays a background, non-dynamical volume element.

The nonlinear action of Weyl transverse gravity takes the following form in the framework of the self-coupling problem:
\begin{equation}
\mathscr{A}:=\frac{1}{\lambda^2}\int\text{d}\mathscr{V} R(\hat{g}),\label{eq:ren1}
\end{equation}
where $R(\hat{g})$ has the same functional form as the Ricci scalar of a metric $\hat{g}_{ab}$ whose determinant is constrained by the condition $\mbox{det}(\hat{g})=\mbox{det}(\eta)$, and $\text{d}\mathscr{V}$ is the flat volume element, which is non-dynamical. For us $\hat{g}_{ab}$ is just a tensor field which lives in a flat background. To make this explicit, let us use the well-known fact \cite{Rosen1940} that one can express the Ricci scalar in terms of the covariant derivatives associated with the flat metric $\eta_{ab}$, denoted by $\nabla$, and integrate by parts to put the action in the form:
\begin{align}
\mathscr{A}=\frac{1}{4\lambda^2}\int\text{d}\mathscr{V}\left[2\hat{g}_{bj}\delta^a_k\delta^i_c-{\hat{g}}^{ai}\hat{g}_{bj}\hat{g}_{ck}\right]\nabla_a{\hat{g}}^{bc}\nabla_i{\hat{g}}^{jk}.\label{eq:ren12b}
\end{align}
Written in this form, it is clear that it describes a special relativistic field theory, albeit nonlinear. This form of the action is the natural one in the framework of the self-coupling problem. Starting with \eqref{eq:ren1} is just convenient to make contact with our knowledge of general relativity but, indeed, one first obtains the action \eqref{eq:ren12b}, which can be optionally supplemented with an irrelevant surface term to give \eqref{eq:ren1}; see \cite{Barcelo2014} for an analogue discussion in the case of general relativity. The covariant notation we are using makes clear that this theory is invariant under general coordinate transformations. Moreover, by construction this action is invariant under transverse diffeomorphisms, whose infinitesimal form is
\begin{equation}
\delta_\xi {\hat{g}}^{ab}:=\mathcal{L}_\xi{\hat{g}}^{ab},\qquad \nabla_a\xi^a=0.\label{eq:diff1}
\end{equation}
The composite object $\hat{g}_{ab}$ is defined as
\begin{equation}
\hat{g}_{ab}:=\kappa^{-1/4}g_{ab}.
\end{equation}
Here $\kappa:=\mbox{det}(g)/\mbox{det}(\eta)$. This definition makes the action automatically invariant under Weyl transformations, whose infinitesimal version is
\begin{equation}
\delta_\omega g_{ab}:=\omega\, g_{ab}.\label{eq:conf1}
\end{equation}
Written in terms of $g_{ab}$ the action is given by:
\begin{align}
\mathscr{A}
=\frac{1}{4\lambda^2}\int\text{d}\mathscr{V}\,\kappa^{1/4}\Big\{[2g_{bj}\delta^a_k\delta^i_c-g^{ai}g_{bj}g_{ck}]\nabla_a g^{bc}\nabla_i g^{jk}+\nonumber\\
+\frac{1}{\kappa}\delta^a_b\delta^i_c\nabla_ag^{bc}\nabla_i\kappa -\frac{1}{2\kappa}g^{ai}g_{bc}\nabla_ag^{bc}\nabla_i \kappa
-\frac{1}{8\kappa^2}g^{ai}\nabla_a \kappa\nabla_i \kappa\Big\}.\label{eq:finaction}
\end{align}
One can see that the transformations \eqref{eq:diff1} and \eqref{eq:conf1} combine in a way that makes this action invariant under transverse diffeomorphisms (now acting on the field $g_{ab}$) as well as Weyl transformations. The action \eqref{eq:finaction} is thus the most general nonlinear covariant action quadratic in the derivatives of the field $g_{ab}$ and satisfying these invariance requisites.

To complete the definition of the theory we need to decide how to include matter. The gravity-matter coupling is not unique, but we can appeal to two principles to resolve its non-uniqueness. The first one is the quantum version of the equivalence principle shown as a consequence of Poincar\'e invariance in \cite{Weinberg1964b}: the coupling of the field which describes gravity to matter and to itself must be governed by the same coupling constant $\lambda$. The second principle rests in the observation that conformal invariance is not a symmetry of matter in our world. A safe way to proceed is then to consider that matter fields are not affected by the Weyl transformations \eqref{eq:conf1} (in the language of \cite{Alvarez2010}, matter fields are inert under Weyl transformations). Both principles imply that the resulting action of matter is obtained by replacing $\eta_{ab}$ with the composite field $\hat{g}_{ab}$. Non-minimal couplings to $\hat{g}_{ab}$ are allowed by construction. Notice that no constraints are imposed on the matter sector.

When expressed in terms of the composite variable $\hat{g}_{ab}$, the nonlinear theory of gravity we are considering here has the form of general relativity plus matter, but it is a theory formulated in flat spacetime and the determinant of the metric is fixed by $\mbox{det}(\hat{g})=\mbox{det}(\eta)$. On these grounds it can be understood as an extension of unimodular gravity which is unconstrained and invariant under general coordinate transformations. By the condition on the determinant, the corresponding (regulated) radiative corrections to the cosmological constant sector (vacuum zero-point energies) in the effective action would have the form
\begin{equation}
\int\text{d}\mathscr{V}E_{\text{vac}}.\label{eq:evac}
\end{equation}
These contributions act as a mere constant shift in the effective action, as we are used to in any special relativistic quantum field theory which does not contain gravity. This does not happen when considering general relativity (as an effective quantum field theory \cite{Donogue1994,Burgess2004}) instead, leading to the cosmological constant problem \cite{Burgess2013,Martin2012}.

Radiative corrections are intrinsically a feature of the semiclassical theory in which matter fields are quantized, so that the form of the matter action is fundamental to address any question concerning them. In Weyl transverse gravity, Weyl invariance dictates a fairly different coupling between gravity and matter fields with respect to the situation in general relativity. Although it should be clear that any argument trying to read the form of these contributions from the form of the classical equations of motion alone is misleading, there still exist some confusion in the community in this respect. A well-known procedure to obtain the form of these radiative corrections leads to a definite answer which settles down the issue of radiative stability unambiguously, as dictated by the form of the symmetries of the gravitational action \cite{Carballo-Rubio2015}.
 
One could expect that deviating from the usual solution to the self-coupling problem would incur in disagreement with experimental facts. On the contrary, the theory constructed here indeed describes gravity in a way which is compatible with all the known experiments in gravitation. The field equations are, by construction, traceless. In particular, in the gauge $\mbox{det}(g)=\mbox{det}(\eta)$ they reduce to the trace-free Einstein equations:
\begin{equation}
R_{ab}-\frac{1}{4}Rg_{ab}=\lambda^2\left(T_{ab}-\frac{1}{4}Tg_{ab}\right).\label{eq:feqs}
\end{equation}
As a consequence, any solution of the Einstein field equations for arbitrary values of the cosmological constant is imported to this theory. In principle one could expect additional solutions in which the Ricci scalar and the trace of the stress-energy momentum tensor are not tied up through a constant but a general function, but this is not the case.

This can be seen by means of the following argument. Given the fact that the canonical stress-energy tensor of gravity and matter is conserved under solutions \cite{Rosen1940}, we recover the Einstein field equations with a phenomenological integration constant, unrelated to zero-point energies of matter, playing the role of a cosmological constant; this is shown in Sec. VI-A of \cite{Barcelo2014}. The arguments above show that this effective cosmological constant is not renormalized by radiative corrections: its value is protected by local symmetries. The resulting field equations include potential energies in the matter sector as, even if gravity is not directly coupled to these terms by construction, they inevitably appear in the definition of the canonical stress-energy tensor. Notice that the conservation of the canonical stress-energy tensor is not an additional equation on its own, but a consequence of the translational symmetry of the theory and the field equations \eqref{eq:feqs}. This is in contrast with the situation in usual formulations of unimodular gravity (see the corresponding discussion in \cite{Barcelo2014} and references therein).

In summary, Weyl transverse gravity arises as a promising candidate for reconciling the classical and quantum-mechanical aspects of the cosmological constant problem. Any theory of emergent gravity which is described by means of this theory in the infrared regime would surpass this problem. On the one hand, all the necessary input to obtain this is deceptively simple: that gravity is mediated by gravitons; on the other hand, the cosmological constant problem may be avoided by the occurrence of a non-dynamical volume element.

\subsection{Causal structure}

One of the most important differences between Newtonian gravity and general relativity is that, in the latter, the causal structure of spacetime is dynamical with its evolution being dictated by the gravitational field equations. It is this departure the one behind the qualitative deviations in the gravitational collapse process \cite{Barcelo2015}.

In Newtonian gravity one can perfectly imagine arbitrarily small balls with 
arbitrarily large masses. One can design non-collision scattering processes 
with arbitrarily small impact parameter. These scatterings are characterized by 
an acceleration phase (falling into the gravitational potential) followed by a 
deceleration phase (climbing the gravitational potential). The entire 
scattering is perfectly time symmetric. By considering the balls as hard 
spheres subject to elastic collisions or non-interacting with each other when coming into superposition (i.e., transparent to each other), one could even make a precise head-on collision. This process will also consist of 
the previous two phases. 

Now, general-relativistic situations in which a lump of matter is climbing a 
gravitational potential can easily be found: one can just think of throwing a 
stone upwards over our heads. One can also think of the following idealized 
thought experiment. Two equal balls with sufficiently small mass-to-radius 
ratio are thrown directly towards each other. Imagine that they are transparent 
to each other (for simplicity one could imagine them to be internally rigid to a first 
approximation and forget about changes in their form). In the case in 
which no horizon forms (i.e., sufficiently low kinetic energy in the 
collision), if one neglects the gravitational wave emission due to the 
encounter, the resulting trajectories will be perfectly time symmetric, again 
with acceleration and deceleration phases. Within standard classical general 
relativity the situation radically changes if a horizon forms. In a collision 
with sufficiently high kinetic energy the two balls in the previous example will 
form a horizon. The time-symmetric deceleration phase will disappear and be
substituted by a black hole remnant. Somewhat surprisingly, although white-hole 
solutions exist mathematically, in supposedly realistic situations one never 
encounters potential-climbing cases beyond the Schwarzschild radius.

The fading into oblivion of the white-hole district is arguably the strongest 
departure of general relativity from Newtonian gravity. There seems to be a 
prejudice against exploring the possibility and implications of really having 
genuine bouncing solutions. As we will see, there are indeed good reasons for 
this prejudice to exist, but also for taking the bold leap of exploring beyond 
it. Our analysis shows that this can be done consistently only under the following assumption: it should exist an underlying causality that is explored only when Planck energies 
are at stake. This background causality should be non-dynamical and trivial in the sense of containing no horizons whatsoever, the simplest 
example one can think of being a Minkowskian structure. Otherwise general-relativistic light cones could not suffer such a dramatic turn. The only possibility apparently left would be that the light cones did not quickly reverse its tilting but only slowly recover their unbent positions before collapse. The geometry would be the one described, for instance, in~\cite{Roman1983}, or more recently in \cite{Rovelli2014}.

Our assumption can be motivated both from particle physics and 
condensed matter physics. On the one hand, our proposal connects naturally with 
Rosen's reformulation of general relativity as a nonlinear theory on a flat 
Minkowski background \cite{Rosen1940}. This reformulation indeed goes further 
than the standard formulation of general relativity in the sense that it is a 
convenient effective framework to describe the switching-off of gravity at high 
energies. Rosen's reformulation can be understood as the long-wavelength limit 
of a nonlinear theory of gravitons (see \cite{Barcelo2014} and references 
therein). It is still an open possibility that an ultraviolet completion of 
such a theory would exhibit asymptotic freedom (as its QCD cousin). On the 
other hand, similar ideas also appear when thinking of gravity as an emergent 
notion in a condensed matter framework~\cite{Barcelo2010a} (see also~\cite{Jannes2009}). The nonlinear 
theory of gravity describes in that case the behavior of collective degrees of 
freedom. There, it is reasonable to think that the first quantum gravitational 
effect is that, above some Planckian energy scale, the collective degrees of 
freedom corresponding to gravity are diluted, leaving a Minkowskian background 
for the matter excitations. 

Let us detail how the switching-off of gravity can be described. The action of 
a theory of this sort should be expressible in terms of a composite field 
$g_{ab}=\eta_{ab}+\lambda h_{ab}$ , where $\eta_{ab}$ is the flat background 
metric, $h_{ab}$ the graviton field and $\lambda$ the coupling constant which 
controls the nonlinearities of the gravitational sector as well as the coupling 
to matter (take Rosen's formulation as an example, see~\cite{Barcelo2014}). For 
any nonzero value of $\lambda$ the field equations of this theory are 
equivalent to the Einstein field equations for a metric $g_{ab}$. Thus in this 
regime one recovers general relativity, with a (generally curved) metric 
$g_{ab}$ controlling the effective causality of the spacetime. However, the 
structure of the theory permits us to consider the limit $\lambda\rightarrow0$, 
in which the nonlinearities disappear and matter is effectively decoupled from 
the graviton field, which evolves separately as a free field. In this limit the 
causality of the spacetime, given by $\eta_{ab}$, is no longer dynamical. The 
two conceptual frameworks above suggest that when high-energy phenomena are 
involved, this limit is indeed reached so that the underlying causality in the 
system, which is Minkowskian with no horizons whatsoever, is unveiled.

\begin{figure}[h]%
\vbox{ \hfil  \includegraphics[width=0.67\textwidth]{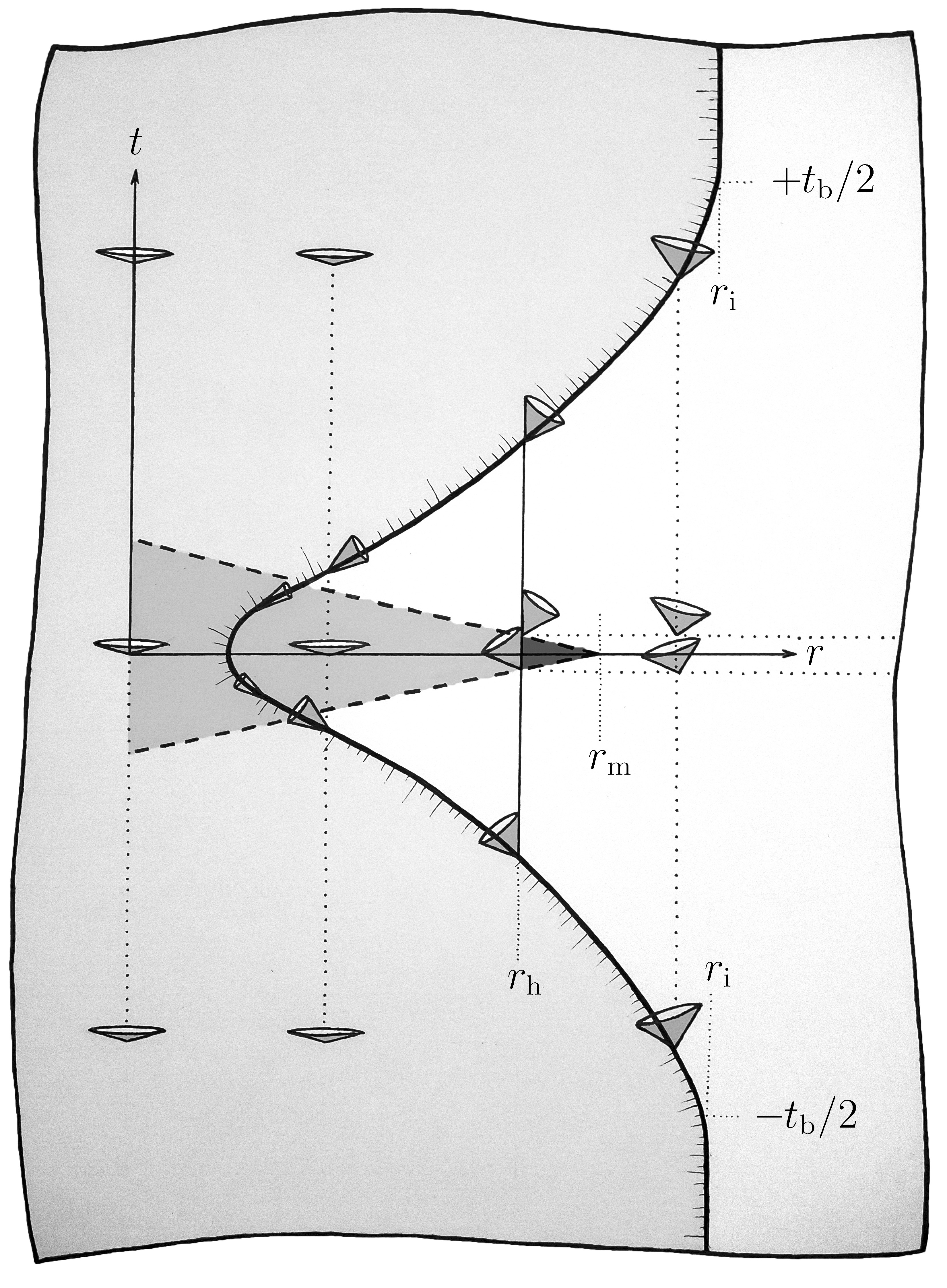}\hfil}
\bigskip%
\caption{The figure represents the collapse and time-symmetric bounce of a 
stellar object in our proposal (the thick line). The past thick dashed line from $r=0$ to $r_{\rm m}$ marks the boundary where the non-standard gravitational effects start to occur. In all the external white region the metric is Schwarzschild. In the region 
between the two thick dashed lines (which extends outside the stellar matter itself) the metric is not 
Schwarzschild, including the small dark gray triangle outside the Schwarzschild 
radius $r_{\rm h}$. The drawing tries to capture the general features of any interpolating geometry. The slope of the almost Minkowskian cones close to the origin has been taken larger than the usual 45 degrees to cope with a convenient and explicit time-symmetric drawing.}
\label{Fig:bh-wh-diagram}%
\end{figure}%

This view prioritizes the role of matter with respect to the dynamical causal 
structure contained in $g_{ab}$, and brings the 
general-relativistic and Newtonian descriptions of matter scattering in the 
presence of gravity a step closer to each other. Let us discuss the qualitative picture of our proposal, thinking about what happens in a local region around the 
distribution of matter undergoing gravitational collapse and making use of the figure \ref{Fig:bh-wh-diagram}. The mathematical details of the solution can be found in \cite{Barcelo2014IJMPD,Barcelo2015}. When a lump of matter undergoes the gravitational collapse process, at some point it will enter the regime in which the local causal structure is 
Minkowskian and there is no trace of gravity. After a scattering process which 
takes place in the absence of gravity and which can be idealized as a first 
approximation as dissipationless, the lump of matter will effectively bounce back, now expanding in 
time. Notice that it is not needed to consider that some kind of repulsive 
force 
acts in this regime, and that its existence would only lead to quantitative 
changes in this picture. If we keep following the expanding distribution of 
matter we will exit the high-energy regime in which the causal structure is not 
dynamical and the usual general-relativistic picture with a gravitational field 
$g_{ab}$ will be restored. However, this dynamical causal structure is a 
secondary character in this story, so that it will naturally adapt itself to 
the 
distribution of matter in spacetime, with the corresponding light cones pointing 
outwards. 

The gravitational switching-off and -on process we have described would have an 
interpretation in general relativity in terms of an effective energy content 
that violates certain energy conditions in some regions. We advance here the following 
interpretation of the ideal dissipationless bouncing process. When the 
collapsing structure reaches Planckian densities it acquires a ``quantum 
modification'' leading to an effective density $\rho-\rho_{\rm q}$. At some 
point this effective density becomes negative and thus repulsive. This negative 
density is compensated by a burst of positive energy that is expelled out of 
the structure through the underlying causality (this is what the bottom thick dashed 
line in Fig.~\ref{Fig:bh-wh-diagram} intends to represent). This positive energy 
burst reaches some point $r_{\rm m}$ outside the gravitational radius. The 
time-reversed process should not be understood as a sort of collapse of this 
energy burst, but rather as describing the attenuation of the effect of its 
propagation through the dynamical geometry.

\begin{figure}[h]%
\vbox{ \hfil  \includegraphics[width=0.35\textwidth]{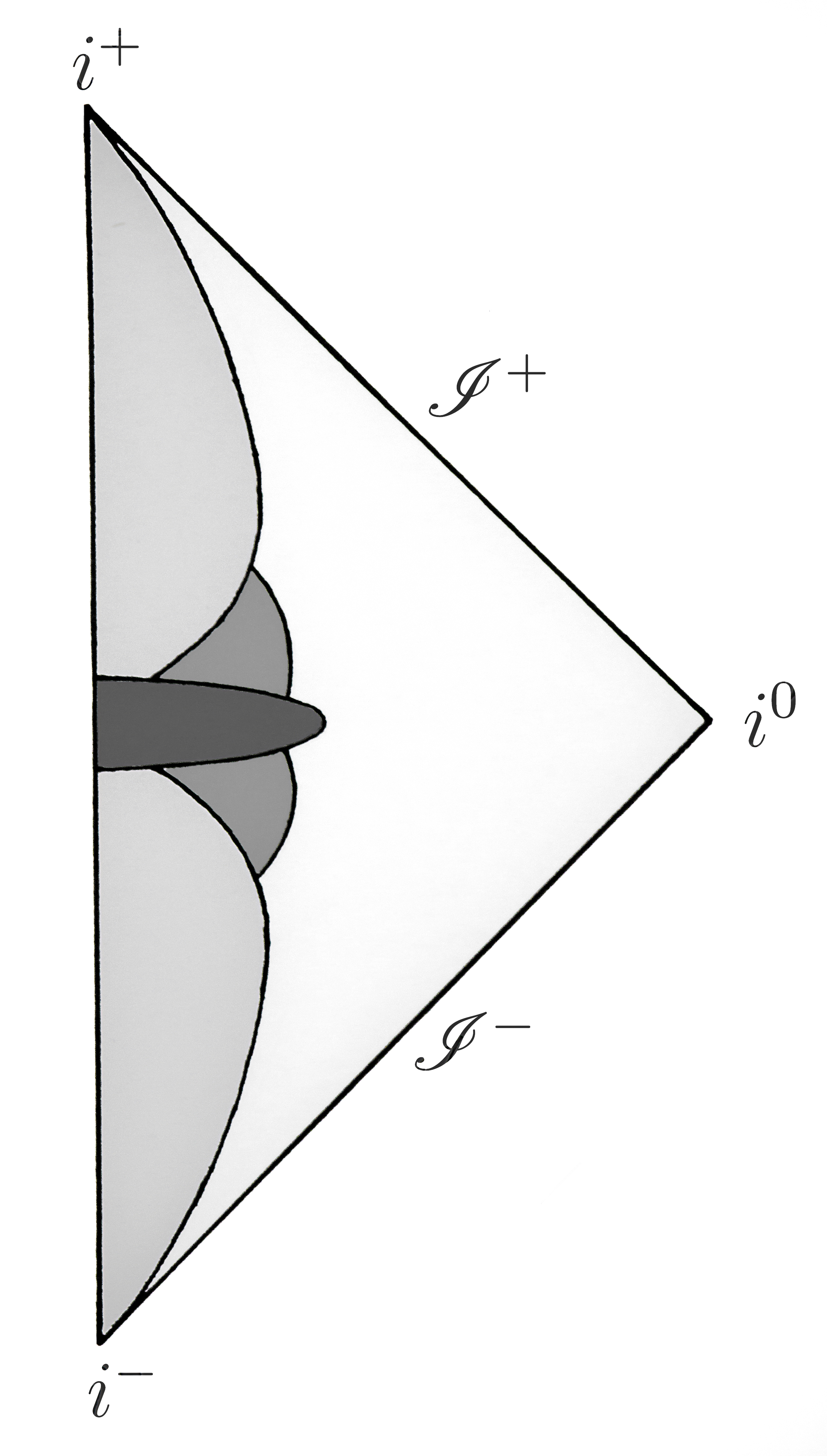}\hfil}
\bigskip%
\caption{The figure represents the Penrose diagram of the proposed geometry. Globally it has the same causality than Minkowski spacetime. Locally it has some peculiarities. The dark gray region represents a non-standard gravitational field, while the down and up gray regions are respectively regions with outer and inner trapped surfaces. The light gray regions on the left-hand side are those filled by matter. There is no long-lived trapping horizons of any sort.}
\label{Fig:penrose-diagram}%
\end{figure}%

In more formal terms, we have shown that there exists an entire family of geometries, with functional and parametric
degrees of freedom which at the moment cannot be fixed without knowing additional features of the underlying theory. However, the most relevant details are robust with respect to these variations, and can be summarized in two points:
\begin{itemize}
\item
The time lapse associated with the collapse is very short (of the order of milliseconds
for neutron-star-like initial configurations) for geometries which simply regularize
the behavior of the collapse near the singularity. This is equally true both for
observers attached to the structure as well as for distant stationary observers.

\item
It is mandatory that a certain open region outside the Schwarzschild radius deviates
from the usual static and spherically symmetric solution. This deviation as well
as the matter bounces will lead to characteristic imprints in the transient phase.
While a detailed study must be done in order to find experimental signatures
of these phenomena, the short timescales associated with all these processes are
encouraging.
\end{itemize}

The plausibility of the time-symmetric bounce process rests on the assumption that the general-relativistic description of gravity is not fundamental, so that modifications to the dynamical behavior of the spatiotemporal causal structure at high energies are conceivable. The standard view is that ``quantum corrections'' to a general-relativistic geometry could only occur in high curvature regions. Our discussion points out that this assertion contains at least one assumption that is typically unstated: The non-existence of a deeper causality of Minkowskian character (that is, with no horizons). If this causality exists, it is not difficult to imagine that the echo of the high-energy collision produced in the location where the classical singularity was supposed to appear would be transmitted outside through the deeper (high-energy) causality, and modify the effective light-cone structure (i.e., the light-cone structure of general relativity) even in places with very small effective curvature. One also has to take into account that the non-general-relativistic modification of the effective light-cone structure we speak about occurs just in a brief transient region, being the manifestation of the rapid process of switching off and on of gravity. The existence of this deeper causality is indeed a necessary and sufficient assumption, thus making a tight connection between certain properties of the high-energy behavior of the gravitational interaction and low-energy solutions which do not present the lifetime problem.

This rapid bounce alternative radically changes the discussion of the possible endpoints of gravitational collapse. In particular, it makes plausible that the final object is a compact object with no horizons whatsoever. As with other compact objects in nature, the structure of the transient would be given by ``dirty'' (non-geometrical) physics whose details still need to be filled out. If this view is indeed realized in nature, black holes could just be  an idealized approximation to the ultimate stationary objects.

The transient phase should leave some traces, for instance, in the physics of gamma ray bursts (GRBs) and its coincident gravitational-wave emission. One would expect some signatures associated with a reverberant collapse. In the collapsar model of GRBs however (see ,e.g., \cite{MacFadyenWoosley1998}) the emission 
zone is supposed to be very far from the collapsed core. This means that the connection between the processes at the core and those at the external wind shells could be very far from direct. The signatures should be clearer however in the spectrum of gravitational waves. The last part of the gravitational-wave signal should clearly distinguish between the standard relaxation towards a black hole and the presence of bouncing processes. 

\section{Conclusions}

There are sensible alternatives to the philosophical assumptions behind the predominant research programs in quantum gravity. In particular, we have considered what we may regard as two of the most rooted principles: that a quantum description of gravity is needed in some regime, and that any gravitational theory containing general relativity should be relational. We have stressed that specific deviations from these alternatives can be compatible with current experimental knowledge, while leading to characteristic experimental signatures which could be possibly tested in future experiments, ranging from mesoscopic physics to astrophysical observations. While work remains to be done in order to understand in detail the possible implications, we have displayed some arguments that point out to generic properties which should be model-independent. Only further research could determine whether these assumptions may stop being philosophical, to become physical.

 
\ack{
Financial support was provided by the Spanish MICINN through Projects No. FIS2011-30145-C03-01 and FIS2011-30145-C03-02 (with FEDER contribution), and by the Junta de Andaluc\'{i}a through Project No. FQM219. R. C-R. acknowledges support from CSIC through the JAE-predoc program, cofunded by FSE.}

\newpage 

\bibliography{c-r_dice2014}	

\end{document}